\documentstyle[12pt,aps]{revtex}

\begin{document}

\hsize = 6.5in
\widetext
\draft
\tighten
\topmargin-48pt

\preprint{EFUAZ FT-97-45-REV}

\title{ON UNUSUAL INTERACTIONS
OF THE $(1/2,0)\oplus (0,1/2)$
PARTICLES\thanks{Submitted to ``Foundation of Physics
Letters".}}

\author{{\bf Valeri V. Dvoeglazov}}

\address{
Escuela de F\'{\i}sica, Universidad Aut\'onoma de Zacatecas \\
Apartado Postal C-580, Zacatecas 98068, ZAC., M\'exico\\
Internet address:  valeri@cantera.reduaz.mx\\
URL: http://cantera.reduaz.mx/\~\,valeri/valeri.htm
}

\date{First version: June 1997. Revised: December 1997}

\maketitle

\begin{abstract}
We continue to study  the `fermion -- 4-vector potential' interactions
in the framework of the McLennan-Case construct which is a
reformulation of the Majorana theory of the neutrino. This theory
is shown after applying Majorana-like {\it anzatzen} to give rise to
appearance of unusual terms as $\bbox{\sigma}\cdot [{\bf A}\times
{\bf A}^\ast]$, which were recently discussed in non-linear optics.
\end{abstract}



\baselineskip24pt

As a result of extracting solid data  certifying  the
existence of the mass of neutrino in the LANL experiment~\cite{APS} the
interest in the Majorana-like models has grown considerably. The
McLennan-Case reformulation~\cite{MLC} of the Majorana theory~\cite{MAJOR}
got further development in the papers of Ahluwalia~\cite{DVA}
and myself~\cite{DVO0,DVO}. In 1996 I received private communications
from D. V. Ahluwalia~\cite{DVAPR} about unusual interactions of neutral
particles in his model which is closely related with the Case
consideration. Even before I learnt about the possible importance
of phase factors of corresponding field functions in defining
the structure of the mass term~\cite{Pashkov}.  They gave initial impulse
in writing this work.  Further investigations from different
standpoints~\cite{Esposito} (compare also with results of non-linear
optics~\cite{OPTICS}) produced simultaneously with this work were also
very incentive in my attempts to solve the problem
{\it rigorously}.\footnote{Obviously, the Evans {\it et al.} derivation
of similar terms [``{\it The Enigmatic Photon. Vol. 3}"  (Kluwer Academic,
Dordrecht, 1996), pp. 9-16, 187-189]  has no any sense in the presented
form.  It should be regarded as completely erroneous until that time when
needed clarifications and corrections would be given .  But, the Esposito
derivation of the term $\sim\bbox{\sigma}\cdot [{\bf A}\times {\bf A}^\ast
]$ is correct. I am grateful to him for sending me the alternative proof
before the publication.}

The main result of the present paper is the theoretical proof of possible
physical significance of the term $\bbox{\sigma}\cdot [{\bf A}\times {\bf
A}^\ast ]$ in the interaction of $(1/2,0)\oplus (0,1/2)$ fermions. In the
process of calculations we use the notation and the metric of ref.~[2b].
The Dirac equation is written
\begin{equation}
(\gamma^\mu \partial_\mu +
\kappa ) \psi =0\, ,\label{1} \end{equation}
where $g^{\mu\nu} = \mbox{diag} ( -1, \, 1,\, 1,\, 1)$
and $\gamma^\mu$ are the Dirac matrices. Their
explicit form can be chosen as follows
\begin{eqnarray} \gamma^0
=\pmatrix{0&-i\cr -i&0\cr}\, , \, \gamma^i = \pmatrix{0&i\sigma^i\cr
-i\sigma^i &0\cr}\, , \, \gamma^5 = \pmatrix{1&0\cr
0& -1\cr} .
\end{eqnarray}
The Pauli charge-conjugation $4\times 4$ matrix is then
\begin{eqnarray}
C = \pmatrix{0&\Theta\cr
-\Theta &0\cr}\, , \, \mbox{where}\,\, \Theta = \pmatrix{0&-1\cr
1&0\cr}\, .
\end{eqnarray}
It has the properties
\begin{mathletters}
\begin{eqnarray}
&&C=C^{^T}\, ,\, C^\ast = C^{-1}\, ,\\
&&C\gamma^\mu C^{-1} = \gamma^{\mu^{\,\ast}}\, ,\, C\gamma^5 C^{-1} =
-\gamma^{5^{\,\ast}}\, .
\end{eqnarray}
\end{mathletters}
As opposed to K. M. Case we introduce the interaction with the
4-vector potential in the beginning and substitute $\partial_\mu
\rightarrow \nabla_\mu = \partial_\mu -i e A_\mu$ in the equation
(\ref{1}). For the sake of generality we assume that the 4-vector
potential is a {\it complex} field what is the extension of this
concept comparing with the usual quantum-field consideration.
(In the classical (quantum) field theory the 4-vector potential
in the coordinate representation is usually considered
to be  {\it pure real} function (functional)).
After introducing projections onto subspaces of the {\it chirality}
quantum number
\begin{equation}
\psi_\pm = {1\over 2} (1\pm \gamma^5) \psi\, , \, \gamma_\pm^\mu ={1\over
2}  (1\pm \gamma^5) \gamma^\mu
\end{equation}
we re-write the equation (\ref{1}) and Eq. (3) of ref.~[2b]
\begin{equation}
(\gamma^\mu \nabla_\mu^\ast +\kappa) C^{-1} \psi^\ast =0\, ,
\end{equation}
which already describe the interactions of (anti) fermion
with the {\it complex} 4-vector potential, to the following set
\begin{mathletters}
\begin{eqnarray}
&&\gamma^\mu_+ \nabla_\mu \psi_- +\kappa \psi_+ =0\, ,\\
&&\gamma^\mu_- \nabla_\mu \psi_+ +\kappa \psi_- =0\, ,\\
&&\gamma^\mu_+ \nabla_\mu^\ast C^{-1} \psi_+^\ast +\kappa
C^{-1}\psi_-^\ast =0\, ,\\
&&\gamma^\mu_- \nabla_\mu^\ast C^{-1} \psi_-^\ast +\kappa
C^{-1}\psi_+^\ast =0\, .
\end{eqnarray}
\end{mathletters}
On using the matrices $\eta^\mu = C\gamma^\mu_-$ and $\eta^{\mu^{\ast}}
= \gamma^\mu_+ C^{-1}$ and $\varphi = \psi_+ = {1\over 2} (1+\gamma^5)
\psi$\,, \, $\chi = C^{-1} \psi_-^\ast$ we obtain
\begin{mathletters}
\begin{eqnarray}
\eta^{\mu^\ast} \nabla_\mu \chi^\ast + \kappa \varphi &=& 0\, ,
\label{eq1} \\
\eta^{\mu} \nabla_\mu \varphi + \kappa \chi^\ast &=& 0\, ,
\label{eq2} \\
\eta^{\mu^\ast} \nabla_\mu^\ast \varphi^\ast + \kappa \chi
&=& 0 \, , \label{eq3} \\
\eta^{\mu} \nabla_\mu^\ast \chi + \kappa \varphi^\ast
&=& 0\, . \label{eq4}
\end{eqnarray} \end{mathletters}
in the sub-space of the {\it positive} chirality quantum number. And
with the matrices $\zeta^\mu = \gamma^\mu_- C^{-1}$, $\zeta^{\mu^{\ast}}
= C\gamma^\mu_+$ and the notation $\eta=\psi_-$, $\xi =
C^{-1}\psi_+^\ast$ we obtain the set
\begin{mathletters}
\begin{eqnarray}
\zeta^{\mu^\ast} \nabla_\mu \eta + \kappa \xi^\ast &=& 0\, ,\label{eq11}\\
\zeta^{\mu} \nabla_\mu \xi^\ast + \kappa \eta &=& 0\, ,\label{eq21}\\
\zeta^{\mu^\ast} \nabla_\mu^\ast \xi + \kappa\eta^\ast &=& 0\,
,\label{eq31} \\
\zeta^{\mu} \nabla_\mu^\ast \eta^\ast + \kappa \xi &=&
0\, ,\label{eq41} \end{eqnarray} \end{mathletters} for the {\it negative}
chirality quantum number.  One can use  four equations of these sets to
describe the physical system. If now apply the Majorana condition given
by Case $\psi_- = C^{-1} \psi_+^\ast$ (see Eq.  (8) in~[2b]) one
can arrive at
$\chi =\varphi$ and
\begin{equation} \nabla_\mu^\ast \varphi \equiv
\nabla_\mu \varphi\,\,\, , \,\,\mbox{hence,}\,\, \, A_\mu = - A_\mu^\ast
\end{equation}
as a consequence of the compatibility condition of the set of equations
(\ref{eq1}-\ref{eq4}). The 4-potential becomes to be {\it pure imaginary}.
This model seems to be perfectly possible after redefining the phase
factor between positive- and negative- energy solutions in the field
operator of the 4-vector potential.

Furthermore, it is difficult to extract the new
physical content from the modification of the Majorana {\it anzatz} such
that $\psi_- = e^{i\alpha(x)} C^{-1} \psi_+^\ast$ and, therefore, $\chi =
e^{-i\alpha (x)} \varphi$.
We come to
\begin{mathletters} \begin{eqnarray}
&&\eta^\mu \nabla_\mu \varphi +\kappa e^{i\alpha (x)} \varphi^\ast = 0\, ,\\
&&\eta^{\mu^{\ast}} \nabla_\mu^\ast \varphi^\ast +\kappa e^{-i\alpha (x)}
\varphi = 0\, ,  \\
&&\mbox{and} \quad \partial_\mu \alpha = e (A_\mu + A^\ast_\mu ) ,
\end{eqnarray}
\end{mathletters}
thus recovering (with $\sigma^{\mu\nu} = {i\over 2} [\gamma^\mu ,
\gamma^\nu ]_-$ )
\begin{equation} \left [\nabla^\mu \nabla_\mu
-i\sigma^{\mu\nu} \nabla_\mu \nabla_\nu -\kappa^2 \right ] \varphi =0\, ,
\end{equation}
and its complex conjugate.

From the above consideration it seems that we failed to derive the needed
term. But, we wish to insist on the general case.  In  order to proceed
let us observe that in the set (\ref{eq1}-\ref{eq4}) and
(\ref{eq11}-\ref{eq41}) the second and the third equations of each set are
complex conjugates each other; the first equation and the fourth equation
are also complex conjugates each other.  If we do not want to introduce
such strong restrictions on the 4-vector potential as above it is logical
to introduce different Majorana-like {\it anzatzen} for these subsets.
This is perfectly possible after one reminds that the subspaces of
different $CP$ quantum number are {\it independent} ones for certain
states.  On this basis, firstly, we re-write the  Dirac equation
and its charge conjugate to another set ($\varphi$ and $\chi$
are two-component spinors):
\begin{mathletters} \begin{eqnarray}
\eta^\mu \nabla_\mu \varphi +\kappa C\varphi &=& 0\, ,\\
\eta^{\mu^\ast} \nabla_\mu^\ast \varphi^\ast +\kappa C^{-1}
\varphi^\ast &=& 0\, ,\\
\eta^\mu \nabla_\mu^\ast \chi +\kappa C\chi &=& 0\, ,\\
\eta^{\mu^\ast} \nabla_\mu \chi^\ast +\kappa C^{-1} \chi^\ast &=& 0\, .
\end{eqnarray}
\end{mathletters}
Next, we set up the following {\it anzatz}
$$C^{-1} \psi_{\pm}^\ast = \mp P \psi_{\pm}\, ,$$
where $P$ is the space inversion operator. Finally,
marking the resulting subsets of equation by some discrete quantum number
(we denote them as ``$s$" and ``$a$") one obtains\footnote{One
could also obtain similar subsets of equations
after the application of the modified Majorana {\it anzatz}
\begin{equation}
\psi_- = \wp_{_{S,A}} C^{-1} \psi_+^\ast\, .
\end{equation}
Here, $\wp_{_{S,A}} =\pm 1$;
the  upper sign  being used for
the first subset, Eqs.  (\ref{eq2},\ref{eq3}) and
the down sign being used
for the second subset, Eqs. (\ref{eq11},\ref{eq41}). This is possible
because the sub-spaces of different chirality quantum numbers can also
be considered as the  independent subspaces and we can choose
any two equations from the both subsets. But, this explanation
can be still considered as obscure by someone due to the discussion in the
first part of the Letter. In my opinion, the proper consideration of the
theory of 4-vector potential is necessary to clarify this point.}
\begin{mathletters} \begin{eqnarray}
\eta^{\mu} \nabla_\mu^\ast
\chi_s +\kappa \chi_s^\ast &=& 0\, ,\\ \eta^{\mu^{\ast}} \nabla_\mu
\chi_s^\ast +\kappa \chi_s &=& 0\, .  \end{eqnarray} \end{mathletters} and
\begin{mathletters} \begin{eqnarray} \eta^\mu \nabla_\mu \varphi_a -\kappa
\varphi_a^\ast &=& 0\, ,\\ \eta^{\mu^\ast} \nabla_\mu^\ast \varphi_a^\ast
-\kappa \varphi_a &=& 0\, .  \end{eqnarray} \end{mathletters} As a result
we obtain second-order equations for $\chi_s$ and $\varphi_a$:
\begin{mathletters} \begin{eqnarray} \left [ \nabla^\mu \nabla_\mu^\ast -
i\sigma^{\mu\nu}\nabla_\mu \nabla_\nu^\ast - \kappa^2 \right ]\chi_s
(x^\mu) &=& 0\, ,\\ \left [ \nabla^{\mu^\ast} \nabla_\mu -
i\sigma^{\mu\nu}\nabla_\mu^\ast \nabla_\nu - \kappa^2 \right ]\varphi_a
(x^\mu) &=& 0\, \end{eqnarray} \end{mathletters} and their complex
conjugates.

One can proceed further with
transformations of these equations to the accustomed forms.
This is only algebraic exercises.
One can see the existence of ``new terms" in the equations:
we proved that some physical states of the spin-1/2 fermion
have interactions of the form
\begin{equation}
i\epsilon^{ijk} \sigma^k \nabla_i  \nabla_j^\ast \rightarrow
+ie^2\bbox{\sigma}\cdot \left [ {\bf A} \times {\bf A}^\ast \right ]\, ,
\end{equation}
for ``$s$" states, and  with the inverse sign, for the ``$a$" states.
(The unit system $c=\hbar =1$ is used.)

At last we note that
the $(1/2,0)\oplus(0,1/2)$ field operator is
naturally decomposed into the parts $\Psi = \psi_{_S} +\psi_{_A}$
\begin{eqnarray}
\psi_{_S} (x^\mu) &=& \int \frac{d^3 {\bf p}}{(2\pi)^3
2E_p} \left \{ \left [ u_{_{\uparrow}} (p^\mu) c_{_{\uparrow}}
(p^\mu) + u_{_{\downarrow}} (p^\mu) d_{_{\downarrow}} (p^\mu) \right ]
e^{-i\phi} +\right .  \nonumber\\
&+& \left . \left [ C u_{_{\uparrow}}^\ast (p^\mu)
c_{_{\uparrow}}^\dagger (p^\mu)
+ C u_{_{\downarrow}}^\ast (p^\mu) d_{_{\downarrow}}^\dagger (p^\mu)
\right ] e^{+i\phi} \right \} \,\, , \\
\psi_{_A} (x^\mu) &=& \int
\frac{d^3 {\bf p}}{(2\pi)^3 2E_p} \left \{ \left [ u_{_{\uparrow}} (p^\mu)
d_{_{\uparrow}} (p^\mu) + u_{_{\downarrow}} (p^\mu)
c_{_{\downarrow}} (p^\mu) \right ]
e^{-i\phi} -\right .  \nonumber\\ &-& \left . \left [ C
u_{_{\uparrow}}^\ast (p^\mu) d_{_{\uparrow}}^\dagger (p^\mu) + C
u_{_{\downarrow}}^\ast (p^\mu) c_{_\downarrow}^\dagger (p^\mu) \right ]
e^{+i\phi} \right \} \,\, , \\
\end{eqnarray}
where $\phi = (Et - {\bf p}\cdot {\bf x})/\hbar$. As easily
demonstrated both parts satisfy (separately  each other) the Dirac
equation.  Certain relations between creation/annihilation operators are
assumed.  They are dictated by the modified Majorana-like {\it anzatzen}.

Finally, I would like to present references to some works which, in
my opinion, would be relevant to further discussions of the questions
put forth here.  Several works already revealed importance of
the term $\bbox{\sigma}\cdot [ {\bf A} \times {\bf A}^\ast ]$ in the
non-linear optics. Other works  are~\cite{Ziino}, where the concept of
{\it two} coordiante-space Dirac equations have been re-discovered
independently (cf.~[6b,d,e] and~\cite{Markov,Belin}); and
ref.~\cite{DVAGR}, where the matters of interface between gravity and
quantum mechanics have been firstly discussed rigorously.

{\it Acknowledgments.}
I acknowledge discussions with Profs. D. V. Ahluwalia, A. E. Chubykalo,
A. Lakhtakia and A. F. Pashkov. I am obliged to Profs. E.
Recami and S.  Esposito for useful information.  In fact, this Letter
is the continuation of their works.

I am grateful to Zacatecas University, M\'exico, for a
professorship.  This work has been partly supported by the Mexican
Sistema Nacional de Investigadores, the Programa de Apoyo a la Carrera
Docente and by the CONACyT, M\'exico under the research project 0270P-E.

\end{document}